\documentclass[10pt,aps,pre,twocolumn,final,letterpaper,superscriptaddress,longbibliography,showkeys]{revtex4-2}
\usepackage[utf8]{inputenc}
\usepackage{amsmath,amsfonts,amssymb}
\usepackage{graphicx}
\usepackage{hyperref}
\usepackage{xcolor}
\usepackage{quotes}
\definecolor{goodblue}{RGB}{0, 91, 187}
\hypersetup{
  colorlinks=true,
  allcolors=goodblue,
  urlcolor=goodblue,
  citecolor=goodblue,
  pdfborder={0 0 0},
  breaklinks=true,
}

\usepackage{setspace}

\renewcommand{\epsilon}{\varepsilon}

\begin{document}

\title{The interplay of network structure and correlated infectious traits in epidemic models} 

\author{Abhay Gupta}
\email{abhay.gupta@virginia.edu}
\affiliation{Department of Biology, University of Virginia, Charlottesville, Virginia 22903, USA}

\author{Nicholas W. Landry}
\email{nicholas.landry@virginia.edu}
\affiliation{Department of Biology, University of Virginia, Charlottesville, Virginia 22903, USA}
\affiliation{School of Data Science, University of Virginia, Charlottesville, Virginia 22903, USA}
\affiliation{Vermont Complex Systems Institute, University of Vermont, Burlington, Vermont 05405, USA}

\begin{abstract}
Individual contributions to the spread of an epidemic vary widely due to an individual's location in a social network and their intrinsic ability to transmit or contract diseases. While the effect of heterogeneous population structure and infection rates is well-understood, less studied is the impact of population-level covariance between susceptibility and transmissibility, despite empirical evidence showing that both susceptibility and transmission vary across individuals.

We introduce a mathematical modeling framework incorporating population subgroups, each with their own joint distribution of susceptibility and transmissibility. We apply this framework to the susceptible-infected-recovered (SIR) model to examine the effect of community structure and degree heterogeneity. We derive analytical expressions for the basic reproduction number, which, when reduced, corroborates prior results and validate these results with numerical simulations. We pair these estimates with simulations exploring first, the temporal dynamics of this model with the homogeneous SIR model, and second, implications for effective social intervention. This analysis provides a foundation for future studies exploring the interplay between structural and dynamical heterogeneity in infectious disease transmission.
\end{abstract}

\maketitle

\raggedbottom
\section{\label{sec:introduction} Introduction}

Individual contributions to disease transmission have long been known to be heterogeneous; although Kermack and Mckendrick's foundational paper in 1927 is best known for the introduction of the susceptible-infected-recovered (SIR) model~\cite{kermack_contribution_1927}, this paper also introduced more general models incorporating heterogeneous transmission across the duration of an infection. While heterogeneous transmission can be driven by many different factors such as environmental variation, multiple pathogenic variants, and cross-reactive immunity, for example, here we frame heterogeneous transmission as an outcome of individual traits and contact structure, similar to the framework introduced in Ref.~\cite{vanderwaal_heterogeneity_2016}. In line with this approach, we model all traits related to social behavior, such as sociality, as a feature of the contact structure of a social system. We refer to non-social traits that directly affect the transmission as \textit{infectious traits} and the collection of these traits as an individual's \textit{infectious profile}.

Examples of infectious traits include an individual's infectious period (also known as the generation interval), viral shedding rate, infection severity, and onset of disease symptoms. For convenience, we combine these traits into \textit{susceptibility}, measuring an individual's likelihood of contracting a disease; \textit{transmissibility}, measuring an individual's ability to transmit a disease; and \textit{recovery rate}, specifying the duration of an infection. Each of these traits has been shown to be quite heterogeneous in empirical contexts.

Heterogeneous susceptibility can be caused by many factors such as prior exposure, individual behaviors like mask-wearing, and heterogeneous immune responses. This variation has been documented in many different systems; for example, in gypsy moths (\textit{Lymantria dispar}), wild populations were found to have more variation in susceptibility compared with lab-reared populations. In a human context, Ref.~\cite{tuschhoff_detecting_2024} quantified heterogeneity in individual susceptibility using contact-tracing data, corroborating other studies demonstrating heterogeneous susceptibility~\cite{dwyer_host_1997, ezenwa_host_2004}.

Heterogeneous transmissibility can be due to factors such as pathogen load, symptom severity, or infection duration. For example, studies show stress can affect the duration of an infection~\cite{glaser_stressinduced_2005} and can indirectly affect an individual's infectiousness by increasing mucosal secretions, vasodilation, and sneezing for example~\cite{cohen_social_1997}. Therefore, variation across stress levels in a population could be enough to generate transmission heterogeneity. Testosterone levels are also correlated with transmissibility; males with high testosterone tend to have less robust immune responses \cite{roberts_testosterone_2009, edler_experimentally_2011}. The distribution of secondary infections, while likely confounded by heterogeneous susceptibility, is nonetheless quite heterogeneous across a range of diseases and outbreaks~\cite{taube_openaccess_2022, althouse_superspreading_2020}, indicating significant transmission heterogeneity.

Heterogeneous individual traits independently impact epidemic dynamics when compared to a population of individuals with homogeneous traits. For example, Ref.~\cite{miller_epidemic_2007} modeled individual susceptibility and infectivity as independent distributions and found that while the reproduction number only depends on the average transmissibility, that the extinction probability of an epidemic increases with the variance of the transmissibility. In later work, Ref.~\cite{miller_spread_2009} demonstrated that while heterogeneity in transmissibility affects the extinction probability, heterogeneity in the susceptibility affects the epidemic final size. These studies assume, however, that first, these distributions are independent, and second, that these traits follow the same distribution for all individuals. The variation in susceptibility and transmissibility need not be independent, however; these traits may be negatively or positively correlated~\cite{hawley_does_2011}. Likewise, subgroups in a population may have different joint distributions of susceptibility and transmissibility.

Until recently, however, relatively few mathematical epidemic models have considered the joint distributions of these traits, much less their distribution across subgroups. Ref.~\cite{harris_infections_2025} introduced an extension of the SIR model where a well-mixed population is endowed with a joint distribution of susceptibility and transmissibility, allowing them to modulate the correlation between those two traits. They found that positive correlations increase the reproduction number, resulting in larger infection peaks and faster spread, while, negative correlations decrease the reproduction number, leading to smaller, delayed infection peaks. Ref.~\cite{tuschhoff_heterogeneity_2025} comprehensively surveyed the literature, finding 9 papers which modeled correlations between susceptibility and transmissibility. In contrast to Ref.~\cite{harris_infections_2025}, this paper also implemented this correlated model as a stochastic process. Neither of these studies, however, consider the interplay between network structure and correlated infectious traits.

Moving beyond this well-mixed assumption is important for translating these theoretical results to empirical contexts for two reasons, first, empirical evidence demonstrates that contact patterns are heterogeneous for both human \cite{mossong_social_2008, cattuto_dynamics_2010, barrat_empirical_2013, pei_optimizing_2021} and animal populations \cite{collier_breathing_2025, sah_unraveling_2017, sah_multispecies_2019}, and second, these contact patterns are determinant of the spread of contagion \cite{brockmann_hidden_2013}. Thus, incorporating network structure is critical for predicting population-level dynamics, identifying critical transmission pathways, and designing effective interventions.

Degree heterogeneity and community structure are two examples of network structure which affect the spread of an epidemic in significant ways. Early work derived the reproduction number, $R_0\approx\beta\langle k^2\rangle/(\gamma\langle k\rangle)$, for the network configuration model~\cite{pastor_satoress_2001}, implying that for degree distributions with unbounded variances, that the reproduction number is unbounded as well. In contrast, while community structure does not, to first approximation, change the reproduction number for equally-sized communities~\cite{landry_opinion_2023}, it does affect the timescale over which an epidemic spreads, creating a bottleneck confining infections to a single community~\cite{huang_epidemic_2007, onnela_structure_2007, karsai_small_2011, wu_how_2008, salathe_dynamics_2010}. However, unequally-sized communities can change the reproduction number because of changes in local network density~\cite{landry_opinion_2023}.

Few studies model both correlations between infectious traits and network structure. Most closely related to our contribution is Ref.~\cite{gou_how_2017} which considered heterogeneous transmission and recovery rates, capturing individual susceptibility and infection duration. From a structural perspective, Ref.~\cite{allard_siam_2023} explores susceptibility and transmissibility as features of a directed contact network; in-degree and out-degree quantify a node's susceptibility and transmissibility respectively.

We attempt to close this gap by introducing a heterogeneous mean-field equation which can not only model group-specific covariance between susceptibility and transmissibility, but arbitrary subgroup-level mixing patterns as well.

Using this framework, we analytically derive $R_0$ for the network configuration model and the two-community stochastic block model with equally and unequally-sized communities and then validate these results with numerical simulations. We then use our analytical calculations to explore impact of both infectious trait correlation and community structure on the reproduction number. Lastly, we analyze both the temporal dynamics of our model as well as the potential for social interventions. Our findings highlight how biological variation and network structure together influence disease spread and provide new insights into development and interpretation of epidemic models.

\begin{figure*}[ht]
    \centering
    \includegraphics[width=\linewidth]{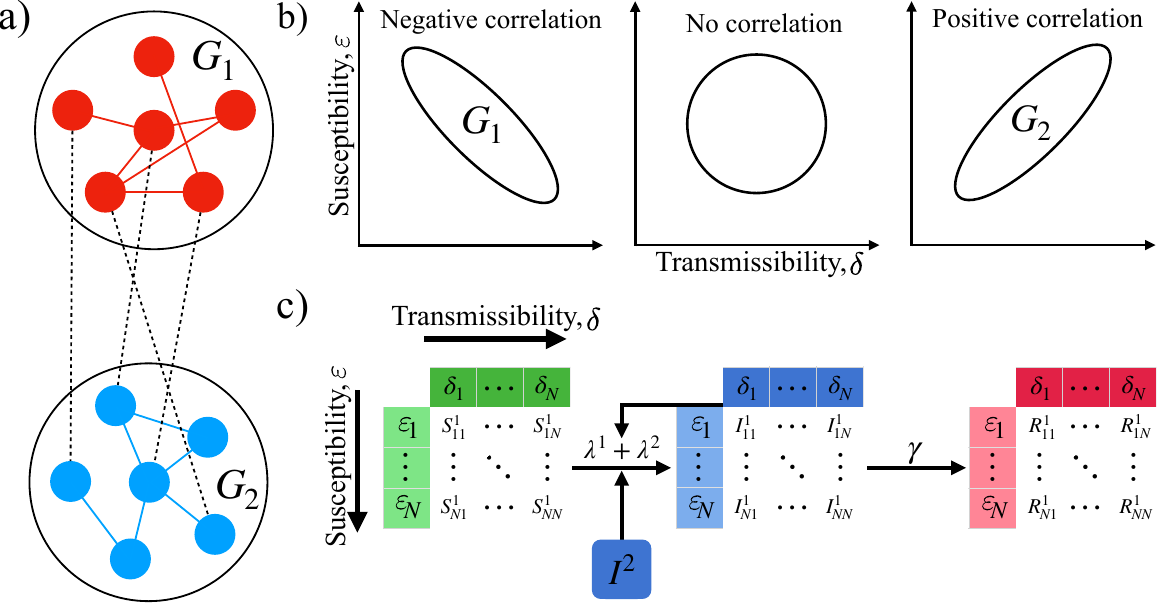}
    \caption{\textbf{An illustration of the $(\delta,\epsilon)$-SIR model.} In panel a), we illustrate our planted partition network model with two communities. Drawing inspiration from Ref.~\cite{harris_infections_2025}, in panel b), we illustrate example correlations between the susceptibility and transmissibility of an individual, illustrating (from left to right) negative correlation, no correlation, and positive correlation between susceptibility and transmissibility, respectively. $G_1$ and $G_2$ denote the joint distributions of transmissibility, $\delta$, and susceptibility, $\epsilon$, for communities 1 and 2 respectively. Lastly, in panel c), we illustrate susceptible, infected, and recovered compartments at each discrete $(\delta_i, \epsilon_j)$ value, denoted for the first community as $S'_{ij}$, $I'_{ij}$, and $R'_{ij}$. $\lambda_1$ and $\lambda_2$ represent the total force of infection from communities 1 and 2, respectively, and $\gamma$ is the recovery rate.}
    \label{fig:framework}
\end{figure*}

\section{\label{sec:mathematicalmodel} Mathematical model}

\subsection{\label{sec:description} Model description}

Here, we describe our mathematical framework. We consider a heterogeneous susceptible-infected-recovered contagion model, which we call the $(\delta, \epsilon)$-SIR model in the following. Similar to Refs.~\cite{harris_infections_2025,tuschhoff_heterogeneity_2025}, we endow an individual $i$ with a susceptibility, $\epsilon_i$, representing $i$'s potential to become infected via an infected contact and a transmissibility, $\delta_i$, representing $i$'s potential to transmit a pathogen to its contacts. 

In the $(\delta,\epsilon)$-SIR model, we define a baseline transmission rate, $\beta$, representing a pathogen's intrinsic infectiousness. Then, the rate at which a susceptible individual $i$ is infected by an infected node $j$ is $\beta f(\delta_j, \epsilon_i)$ where $f$ is an arbitrary real function, mapping to non-negative numbers~\cite{miller_epidemic_2007}. Here, we assume the effects are independent, yielding a transmission rate of $\beta \epsilon_i \delta_j$. This can be equivalently represented as an SIR model on a weighted, directed network where $w_{ij}=\epsilon_i \delta_j$ and $w_{ij}$ may not equal $w_{ji}$. $\epsilon$ and $\delta$ can covary and are jointly distributed in a subpopulation $g$ according to the joint distribution $G^{g}$. While a subpopulation is arbitrary and may comprise on one extreme, the entire population, and on the other extreme, a single node, in this study, we consider subpopulations representing communities and nodes of equal degree. Here, population structure is modeled as a static, undirected network, where nodes represent individuals and unweighted edges represent social contacts.

We consider two models of network structure: the balanced and imbalanced planted partition model \cite{landry_opinion_2023} representing community structure and the network configuration model \cite{fosdick_configuring_2018} representing degree heterogeneity. We employ a heterogeneous mean-field approach similar to those used in \cite{landry_opinion_2023, pastor_satoress_2001, restrepo_approximating_2005}, where we partition nodes into subgroups based on their structural properties, e.g., community or degree.

Here, we describe the differential equations governing the evolution of the distributions of $S$, $I$, and $R$. In contrast to Ref.~\cite{harris_infections_2025} which enforced $\int_{\delta, \epsilon} S(\delta, \epsilon) \, d\delta \, d\epsilon + \int_{\delta, \epsilon} I(\delta, \epsilon) \, d\delta \, d\epsilon + \int_{\delta, \epsilon} R(\delta, \epsilon) \, d\delta \, d\epsilon = 1$, we instead require that $S(\delta, \epsilon) + I(\delta, \epsilon) + R(\delta, \epsilon) = 1$, for every $\epsilon$ and $\delta$. This has the advantage of decoupling the population-wide distribution of $\delta$ and $\epsilon$ from the distribution of individuals in each disease state. One can recover the distributions in Ref.~\cite{harris_infections_2025} from the product of $G(\delta, \epsilon)$ with $S(\delta, \epsilon)$, $I(\delta, \epsilon)$, and $R(\delta, \epsilon)$.

Suppose that a population with $N$ individuals can be decomposed into subgroups $g_1, \dots, g_M$, each of which has a joint distribution of $\delta$ and $\epsilon$ specified by $G^{g}(\delta,\epsilon)$. We denote the fraction of nodes in group $g$ as $P(g)$ and the probability that node in subgroup $g_i$ is connected to a node in subgroup $g_j$ as $P(g_i, g_j)$. Then a mean-field description of the SIR model with heterogeneous susceptibility and transmissibility can be written as
\begin{align}
    \frac{dS^{g}(\delta, \epsilon)}{dt} = &-\beta \epsilon (1-S^{g}(\delta, \epsilon)) \sum_{g'=1}^M P(g\mid g') N(g') \nonumber\\ 
    &\times \int_{\delta', \epsilon'} \delta' G^{g'}(\delta', \epsilon') I^{g'}(\delta', \epsilon') \, d\delta \, d\epsilon, \label{eq:gen-eq1} \\
    \frac{dI^{g}(\delta, \epsilon)}{dt} = &\,\beta \epsilon (1-S^{g}(\delta, \epsilon)) \sum_{g'=1}^M P(g\mid g') N(g') \nonumber\\ 
    &\times\int_{\delta', \epsilon'} \delta' G^{g'}(\delta', \epsilon') I^{g'}(\delta', \epsilon') \, d\delta \, d\epsilon \nonumber\\
    &- \gamma I^{g}(\delta, \epsilon), \label{eq:gen-eq2} \\
    \frac{dR^{g}(\delta, \epsilon)}{dt} = &\,\gamma I^{g}(\delta, \epsilon), \label{eq:gen-eq3}
\end{align}
where $N(g) = NP(g)$ is the number of individuals in subgroup $g$. We discretize $\delta$ and $\epsilon$ as $\delta_1, \dots, \delta_L$ and $\epsilon_1, \dots, \epsilon_L$ and rewrite the integral as a sum to simplify future analysis. Then,

\begin{align}
    \frac{dS^{g}(\delta_i, \epsilon_j)}{dt} \equiv \frac{dS^g_{ij}}{dt} = &-\beta \epsilon_j S^g_{ij} \sum_{g'=1}^M P(g\mid g') N(g') \nonumber\\ 
    &\times \sum_{k=1}^L \sum_{l=1}^L \delta_k G^{g'}_{kl} I^{g'}_{kl} \, \Delta\delta \, \Delta\epsilon, \label{eq:simp-eq1} \\
    \frac{dI^{g}(\delta_i, \epsilon_j)}{dt} \equiv \frac{dI^g_{ij}}{dt} = & \, \beta \epsilon_j S^g_{ij} \sum_{g'=1}^M P(g\mid g') N(g') \nonumber\\ 
    &\times \sum_{k=1}^L \sum_{l=1}^L \delta_k G^{g'}_{kl} I^{g'}_{kl} \, \Delta\delta \, \Delta\epsilon \nonumber\\&-\gamma I^g_{ij}, \label{eq:simp-eq2} \\
    \frac{dR^{g}(\delta, \epsilon)}{dt} \equiv \frac{dR^g_{ij}}{dt} = & \, \gamma I^g_{ij}, \label{eq:simp-eq3}
\end{align}

\noindent where $\Delta\delta=[\max(\delta)-\min(\delta)]/L$ and $\Delta\epsilon=[\max(\epsilon)-\min(\epsilon)]/L$. Note that as $L \to \infty$, we recover the original system of equations in Eqs.~\eqref{eq:gen-eq1}-\eqref{eq:gen-eq3}.

In the following, we first examine the Erd\H os-R\'enyi model to introduce our analytical framework and to compare our results with prior estimates of $R_0$ in Refs.~\cite{harris_infections_2025,tuschhoff_heterogeneity_2025}. We then use this framework to analyze the dynamics of this model on networks with community structure and heterogeneous degree distributions.

This system of equations always has the equilibrium $I^g_{ij}=0$ for all $i$, $j$, and $g$. Most typically, we are interested in the system at the beginning of an epidemic, i.e., when $S^g_{ij}\approx1$ and $R^g_{ij}\approx0$ for every $i$, $j$, and $g$, though every combination where $S^g_{ij} + R^g_{ij} = 1$ is also an equilibrium. Because the equations for $I$ contain all of the terms in both the $S$ and $R$ equations, we focus on the equations for $I$ using the approximations $S^g_{ij}\approx1$ and $R^g_{ij}\approx0$. We consider small perturbations $\tilde{\mathbf{I}}=\mathbf{I}-\mathbf{I}^*$ about the $\mathbf{0}$ equilibrium and Taylor expand the equation
\begin{align}
     F(I^g_{ij}) = &-\gamma I^g_{ij} + \beta \epsilon_j \sum_{g'}P(g\mid g')N(g') \nonumber\\&\times\sum_{k,l}\delta_k G^{g'}_{kl}I^{g'}_{kl} \, \Delta\delta\Delta\epsilon, \label{eq:funcI}
\end{align} 
obtaining
\begin{align}
    \frac{d\tilde{\mathbf{I}}}{dt} = J(\mathbf{0})\tilde{\mathbf{I}}, \label{eq:Reduc-Jacobian}
\end{align}
where $J$ is the Jacobian of $F$. We denote the eigenvalues of $\mathbf{J}$ as $\lambda_i$ and the spectral abscissa as $\nu = \max[\text{Re}(\lambda_i)]$. When $\nu = 0$, this equilibrium becomes unstable, corresponding to the basic reproduction number, $R_0$, equaling $1$. In the following, we analytically derive the eigenvalues of $\mathbf{J}$ for several common network models to obtain an analytical expression for $R_0$.

\subsection{\label{sec:erdos-renyi} Random mixing}

Here, we assume a single subpopulation, i.e., $M=1$, where two nodes $i$ and $j$ connect with probability $p$ which specifies the Erd\H os-R\'enyi model. For a network of size $N$ with a mean degree of $\langle k \rangle$, $2p\binom{N}{2} = \langle k \rangle N$, and so $p\approx\langle k\rangle/N$. Similarly, $p(1,1) = p = \langle k \rangle / N$, $N(1)=1$, we denote $G_{ij} \equiv G^1_{ij}$ for simplicity, and our system of equations reduces to
\begin{align}
    \frac{dS_{ij}}{dt} &= -\beta \epsilon_j S_{ij} \langle k \rangle \sum_{k,l} \delta_k G_{kl} I_{kl} \, \Delta\delta \Delta\epsilon, \label{eq:ER-gen-eq1}\\
    \frac{dI_{ij}}{dt} &= \beta \epsilon_j S_{ij} \langle k \rangle \sum_{k,l} \delta_k G_{kl} I_{kl} \, \Delta\delta \Delta\epsilon - \gamma I_{ij}, \label{eq:ER-gen-eq2}\\
    \frac{dR_{ij}}{dt} &= \gamma I_{ij}.\label{eq:ER-gen-eq3}
\end{align}
Using the approximation $S_{ij}\approx1$,
\begin{align}
    F_{ij}(I_{ij}) = \beta \epsilon_j \langle k\rangle\sum_{k,l} \delta_k G_{kl} I_{kl} \, \Delta\delta \Delta\epsilon -\gamma I_{ij}, \label{ER-funcI}
\end{align}
and the first-order Taylor approximation at $\mathbf{I^*} = \mathbf{0}$ is 
\begin{align}
    \frac{d\tilde{I}_{ij}}{dt} \approx \sum_{k,l} J_{ijkl} \tilde{I}_{kl}, \label{eq:ER-approx}
\end{align}
where the Jacobian, $\mathbf{J}$, is the 4-tensor
\begin{align}
    J_{ijkl}=\frac{dF_{ij}}{dI_{kl}}=-\gamma\mathbf{1}_{i=k,j=l} + \beta\epsilon_j\langle k\rangle G_{kl}\delta_l \, \Delta\delta\Delta\epsilon, \label{eq:ER-jacobian}
\end{align}
where $\mathbf{1}$ is the indicator function.

We vectorize Eqs.~\eqref{eq:ER-approx} and \eqref{eq:ER-jacobian} such that the $(i, j) \text{th}$ index becomes the $\bigl[(L-1)i+j\bigr]\text{th}$ index and expand $I_{ij}$ into $\mathbf{I} = [I_{11}, \dots, I_{1L}, \dots, I_{L1}, \dots, I_{LL}]^T$. Similarly, we expand $\mathbf{J}$ into a $L^2\times L^2$ matrix by mapping indices $(i, j)$ to the $\bigl[(L-1)i+j\bigr]\text{th}$ row and indices $(k, l)$ to the $\bigl[(L-1)k+l\bigr]\text{th}$ column. Then,
\begin{align}
    \mathbf{J} = -\gamma \mathbb{I}^{L^2} + \beta \langle k \rangle \Delta\delta \Delta\epsilon \mathbf{x}\mathbf{y}^T, \label{eq:ER-2d_jacobian}
\end{align}
where $\mathbb{I}^{L^2}$ is the $L^2\times L^2$ identity matrix, 
\begin{equation*}
    \mathbf{x}=[\epsilon_1,\dots,\epsilon_1,\dots,\epsilon_L,\dots,\epsilon_L]^T,
\end{equation*}
and 

\begin{equation*}
    \mathbf{y}=[G_{11}\delta_1,\dots,G_{1L}\delta_1,\dots,G_{L1}\delta_L,\dots,G_{LL}\delta_L]^T.
\end{equation*}

Because the second term of Eq.~\eqref{eq:ER-2d_jacobian} is a rank-1 matrix, the largest eigenvalue of the Jacobian is $\beta \langle k\rangle \mathbf{x}^T\mathbf{y}-\gamma=\beta\langle k \rangle \sum_{i,j} \delta_i G_{ij}\epsilon_j\,\Delta\delta\Delta\epsilon=\beta\langle k\rangle\langle\delta\epsilon\rangle-\gamma$.

When this eigenvalue becomes positive, $\mathbf{0}$ becomes an unstable fixed point, which corresponds to $R_0>1$. Then
\begin{align}
    R_0^{(\text{ER})} = \frac{\beta\langle k\rangle \langle\delta\epsilon\rangle}{\gamma},
\end{align}
which corroborates previous results in Ref.~\cite{harris_infections_2025} after defining $\tilde{\beta} = \beta\langle k\rangle$ and subtracting and adding $\langle\delta\rangle\langle\epsilon\rangle$.

\subsection{\label{sec:sbm} Community structure}

Now that our analytical framework has been described, we turn our attention to more complex network models. The stochastic block model (SBM) is characterized by node labels, $g_i$, specifying the community to which node $i$ belongs and an affinity matrix, $\omega_{g_i, g_j}$, specifying the probability that node $i$ in group $g_i$ is connected to node $j$ in group $g_j$.

We consider a network with two communities and a fixed global mean degree, $\langle k\rangle$. We first analyze the case where the two communities are of equal sizes, and then generalize to imbalanced community sizes.

\subsubsection{\label{sec:balanced_sbm} Balanced communities}

Here, we use our mathematical framework to model two subgroups of equal size. In this case,
\begin{align*}
N(g_1)&=N(g_2)=N/2,\\
P(g_1\mid g_1)&=P(g_2\mid g_2)=p_{\text{in}},\\
P(g_1\mid g_2)&=P(g_2\mid g_1)=p_{\text{out}}.
\end{align*}

Following the analysis of Ref.~\cite{landry_opinion_2023}, $p_{\text{in}}\approx \langle k\rangle(1+e)/N$ and $p_{\text{out}}\approx \langle k\rangle(1-e)/N$, where $e$ is the imbalance parameter, formerly denoted by $\epsilon$ in Ref.~\cite{landry_opinion_2023}. Then, without loss of generality, we write the resulting system of equations for community 1, noting that we can obtain the system of equations by substituting the quantities corresponding to community 1 with those corresponding to community 2 and vice-versa.

\begin{align}
    \frac{dS^1_{ij}}{dt}=&-\beta\epsilon_jS^1_{ij}\Biggl(\frac{\langle k\rangle(1+e)}{2}\sum_{k,l}\delta_k G^1_{kl}I^1_{kl} \nonumber\\ &+ \frac{\langle k\rangle(1-e)}{2}\sum_{k,l}\delta_k G^2_{kl}I^2_{kl} \Biggr),\label{eq:BC-gen1}\\ 
    \frac{dI^1_{ij}}{dt}=&\, \beta\epsilon_jS^1_{ij}\Biggl(\frac{\langle k\rangle(1+e)}{2}\sum_{k,l}\delta_k G^1_{kl}I^1_{kl} \nonumber\\ &+ \frac{\langle k\rangle(1-e)}{2}\sum_{k,l}\delta_k G^2_{kl}I^2_{kl} \Biggr) - \gamma I^1_{ij},\label{eq:BC-gen2}\\
    \frac{dR^1_{ij}}{dt}=&\,\gamma I^1_{ij}.\label{eq:BC-gen3}
\end{align}

Communities 1 and 2 each have their own joint distribution over susceptibility and transmissibility specified by $G^{1}(\delta,\epsilon)$ and $G^{2}(\delta, \epsilon)$, respectively. One might imagine that each community might have a significantly different distribution, where, in one community, susceptibility and transmissibility are highly correlated, whereas in another community, susceptibility and transmissibility might be uncorrelated or even anti-correlated.

Here, we assume that $G^1_{ij}$ and $G^2_{ij}$ not only have the same support over $\delta$ and $\epsilon$, but that the discretization is the same as well, i.e., $L_1=L_2=L$.

We vectorize Eqs.~\eqref{eq:BC-gen1}-\eqref{eq:BC-gen3} by mapping indices $(g, i, j)$, where $g$ is the community label to the index $L^2(g-1)+L(i-1)+j$. Then, $\mathbf{I}=\begin{bmatrix}\mathbf{\tilde{I}_1} \\ \mathbf{\tilde{I}_2}\end{bmatrix}$, where $\mathbf{\tilde{I}}$ is the vectorized $I$ vector described in section \ref{sec:erdos-renyi}, and the Jacobian can be described in block form as 

\begin{align}
    \mathbf{J}&=\beta\langle k\rangle\begin{bmatrix}B_{11} & B_{12} \\ B_{21} & B_{22}\end{bmatrix} - \gamma \mathbb{I}, \\
    B_{ij} &= c_{ij}\mathbf{xy}_i^T,
\end{align}
where $c_{11}=c_{22}=(1+e)\Delta\delta\Delta\epsilon/2$, $c_{12}=c_{21}=(1-e)\Delta\delta\Delta\epsilon/2$, $\mathbf{x}$ is described as before, and 
\begin{equation*}
    \mathbf{y}_k=[\delta_1G^k_{11},\dots,\delta_1G^k_{1L},\dots,\delta_LG^k_{L1},\dots,\delta_LG^k_{LL}]^T.
\end{equation*}
We can exploit this block structure to write $B$ as the sum of two rank-1 matrices namely,
\begin{align}
    B=\mathbf{\hat{x}_1\hat{y}_1}^T+\mathbf{\hat{x}_2\hat{y}_2}^T,
\end{align}
where, for simplicity, we denote 
\begin{equation*}
\mathbf{\hat{x}_1}=\begin{bmatrix}c_{11}\mathbf{x} \\ c_{12}\mathbf{x}\end{bmatrix}\!,\,\, \mathbf{\hat{x}_2}=\begin{bmatrix}c_{21}\mathbf{x} \\ c_{22}\mathbf{x}\end{bmatrix}\!,\,\, \mathbf{\hat{y}_1}=\begin{bmatrix}\mathbf{y}_1 \\ \mathbf{0}\end{bmatrix}\!,\,\,\mathbf{\hat{y}_2}=\begin{bmatrix}\mathbf{0} \\ \mathbf{y}_2\end{bmatrix}\!.
\end{equation*}
Because this is a rank-2 matrix, every eigenvector can be expressed as a linear combination of basis vectors, i.e.,
\begin{align}
    \mathbf{v} = a\mathbf{\hat{x}_1} + b\mathbf{\hat{x}_2}.
\end{align}
Then, the eigenvalue equation becomes
\begin{align}
    \bigl(\mathbf{\hat{x}_1\hat{y}_1}^T+\mathbf{\hat{x}_2\hat{y}_2}^T - \lambda \mathbb{I}\bigr)\bigl(a\mathbf{\hat{x}_1}+b\mathbf{\hat{x}_2}\bigr)=0.
\end{align}
We expand and group by $\mathbf{\hat{x}_1}$ and $\mathbf{\hat{x}_2}$, yielding $\bigl[(\mathbf{\hat{y}_1}^T\mathbf{\hat{x}_1})a+(\mathbf{\hat{y}_1}^T\mathbf{\hat{x}_2})b-\lambda a\bigr]\mathbf{\hat{x}_1}+\bigl[(\mathbf{\hat{y}_2}^T\mathbf{\hat{x}_1})a+(\mathbf{\hat{y}_2}^T\mathbf{\hat{x}_2})b-\lambda b\bigr]\mathbf{\hat{x}_2}=0$.

Because $\mathbf{\hat{x}_1}$ and $\mathbf{\hat{x}_2}$ are linearly independent, both coefficients must be zero, yielding the reduced eigenvalue equation
\begin{align}
    \begin{bmatrix}
        \mathbf{\hat{y}_1}^T\mathbf{\hat{x}_1}-\lambda & \mathbf{\hat{y}_1}^T\mathbf{\hat{x}_2} \\
        \mathbf{\hat{y}_2}^T\mathbf{\hat{x}_1} & \mathbf{\hat{y}_2}^T\mathbf{\hat{x}_2} - \lambda
    \end{bmatrix}\begin{bmatrix}
        a \\ b
    \end{bmatrix} = \begin{bmatrix}
        0 \\ 0
    \end{bmatrix}.
\end{align}
Setting the determinant of this matrix equal to zero, we obtain 
\begin{align*}
    \lambda^2 - &\bigl[(\mathbf{\hat{y}_1}^T \mathbf{\hat{x}_1})+(\mathbf{\hat{y}_2}^T\mathbf{\hat{x}_2})\bigr]\lambda \nonumber\\+&\bigl[(\mathbf{\hat{y}_1}^T\mathbf{\hat{x}_1})(\mathbf{\hat{y}_2}^T\mathbf{\hat{x}_2})-(\mathbf{\hat{y}_2}^T\mathbf{\hat{x}_1})(\mathbf{\hat{y}_1}^T\mathbf{\hat{x}_2})\bigr]=0.
\end{align*}
Solving this quadratic equation and substituting the definition of $\mathbf{\hat{x}}$ and $\mathbf{\hat{y}}$, the eigenvalues are 
\begin{align}
    \lambda_B = \frac{(1+e)}{4}\biggl[\langle \delta\epsilon\rangle_1 + \langle \delta\epsilon\rangle_2 \biggr] \pm \frac{\sqrt{\Delta}}{4},    
\end{align}
where $\Delta=(1+e)^2\bigl[\langle\delta\epsilon\rangle_1-\langle\delta\epsilon\rangle_2\bigr]^2+4(1-e)^2\langle\delta\epsilon\rangle_1\langle\delta\epsilon\rangle_2$.

\noindent Then the largest eigenvalue of the Jacobian is $\lambda=\beta\langle k\rangle\max(\lambda_B)-\gamma$ and the basic reproduction number is
\begin{align}
    R_0=\frac{\beta\langle k\rangle }{4\gamma}\biggl[(1+e)\bigl[\langle\delta\epsilon\rangle_1+\langle\delta\epsilon\rangle_2\bigr]+\sqrt{\Delta}\biggr].
\end{align}

When $\langle \delta\epsilon\rangle_1 = \langle \delta\epsilon\rangle_2$, we recover the expression we obtained for $R_0^{(ER)}$ when we define $\langle\delta\epsilon\rangle=\big[\langle\delta\epsilon\rangle_1 + \langle\delta\epsilon\rangle_2\big]/2$. When the two communities are fully isolated, i.e., $e=1$, $R_0$ depends only on the more positively correlated community; i.e., $R_0 = \beta\langle k\rangle\text{max}\left(\langle\delta\epsilon\rangle_1, \langle\delta\epsilon\rangle_2\right)/\gamma$. A bipartite network, where nodes of types $1$ and $2$ represent communities $1$ and $2$ respectively, corresponds to $e=-1$. In this case, the reproduction number is the geometric mean of the two reproduction numbers one obtains from looking at each community in isolation, i.e., $R_0=\beta\langle k\rangle\sqrt{\langle\delta\epsilon\rangle_1\langle\delta\epsilon\rangle_2}/\gamma$.

\subsubsection{\label{sec:imbalanced_sbm} Imbalanced communities}

Following the analysis for the case of balanced communities, we determine $P(g_i\mid g_j)$ and $N(g_j)$ when the two communities are different sizes. We specify that $N(g_1) = rN$, and $N(g_2) = (1-r)N$ (where $r$ is the fraction of individuals in community 1). As before, we assume that $P(g_1\mid g_1) = P(g_2\mid g_2) = p_{in}$ and $P(g_1\mid g_2) = P(g_2\mid g_1) = p_{out}$.

Following the analysis of Ref.~\cite{landry_opinion_2023} for imbalanced communities, $p_{\text{in}}\approx \langle k\rangle(1+\alpha e) / N$ and $p_{\text{out}}\approx \langle k\rangle(1-e) / N$, where
\[\alpha=\frac{1-(r^2 + (1-r)^2)}{(r^2 + (1-r)^2)}.\] 

\noindent As with the balanced community case, we write the system of equations for community 1 as

\begin{align}
    \frac{dS^1_{ij}}{dt}=&-\beta\epsilon_jS^1_{ij}\Biggl(\langle k\rangle(1+\alpha e)r\sum_{k,l}\delta_k G^1_{kl}I^1_{kl} \nonumber\\ &+ \langle k\rangle(1-e)(1-r)\sum_{k,l}\delta_k G^2_{kl}I^2_{kl} \Biggr),\label{eq:IBC-gen1}\\ 
    \frac{dI^1_{ij}}{dt}=&\,\beta\epsilon_jS^1_{ij}\Biggl(\langle k\rangle(1+\alpha e)r\sum_{k,l}\delta_k G^1_{kl}I^1_{kl} \nonumber\\ &+ \langle k\rangle(1-e)(1-r)\sum_{k,l}\delta_k G^2_{kl}I^2_{kl} \Biggr) - \gamma I^1_{ij},\label{eq:IBC-gen2}\\
    \frac{dR^1_{ij}}{dt}=&\,\gamma I^1_{ij}.\label{eq:IBC-gen3}
\end{align}

We vectorize Eqs.~\eqref{eq:IBC-gen1}-\eqref{eq:IBC-gen3} using the same method described in section \ref{sec:balanced_sbm}. The key difference, however, is that now 
\begin{align*}
    c_{11}&=(1+\alpha e)r\Delta\delta\Delta\epsilon, \\c_{22}&=(1+\alpha e)(1-r)\Delta\delta\Delta\epsilon, \\c_{12}&=(1-e)(1-r)\Delta\delta\Delta\epsilon, \\c_{21}&=(1-e)r\Delta\delta\Delta\epsilon.
\end{align*}
We perform the same block reduction, again obtaining $\bigl(\mathbf{\hat{x}_1\hat{y}_1}^T+\mathbf{\hat{x}_2\hat{y}_2}^T - \lambda \mathbb{I}\bigr)\bigl(a\mathbf{\hat{x}_1}+b\mathbf{\hat{x}_2}\bigr)=0$.

However, now when we substitute the definitions of $\mathbf{\hat{x}}$ and $\mathbf{\hat{y}}$ into the solutions of the eigenvalue equation, we obtain
\begin{align}
    \lambda_B = \frac{(1+\alpha e)}{2}\biggl[r\langle \delta\epsilon\rangle_1 + (1-r)\langle \delta\epsilon\rangle_2 \biggr] \pm \frac{\sqrt{\Delta}}{2},    
\end{align}
where 
\begin{align*}
    \Delta=&\,(1+\alpha e)^2\bigl[r\langle\delta\epsilon\rangle_1-(1-r)\langle\delta\epsilon\rangle_2\bigr]^2\\
    &+4(1-e)^2r(1-r)\langle\delta\epsilon\rangle_1\langle\delta\epsilon\rangle_2.
\end{align*}
Then the basic reproduction number is
\begin{align}
        R_0=&\, \frac{\beta\langle k\rangle }{2\gamma}\biggl[(1+\alpha e)\bigl[r\langle\delta\epsilon\rangle_1\nonumber\\&+(1-r)\langle\delta\epsilon\rangle_2\bigr]+\sqrt{\Delta}\biggr].
\end{align}

\subsection{\label{sec:config-model}Degree heterogeneity}
Our analysis up to now has focused on relatively homogeneous subpopulations, and we have not yet examined the effect of heterogeneity in the number of connections. We represent this structure using the network configuration model, which is specified by a degree distribution, $P(k) = N(k)/N$, and assumes random connection between two nodes proportional to degree, i.e., $P(k\mid k') = kk'/(\langle k\rangle N)$. In this case, we assume that the number of unique degrees is $M$. Then
\begin{align}
    \frac{dS^k_{ij}}{dt} = &-\beta\epsilon_j S^k_{ij}\frac{k}{\langle k\rangle}\sum_{k'}k'P(k')\sum_{l,m} G^{k'}_{lm}\delta_l I^{k'}_{lm},\\
    \frac{dI^k_{ij}}{dt} = &-\gamma I^k_{ij} + \beta \epsilon_j S^k_{ij}\frac{k}{\langle k\rangle}\sum_{k'}k'P(k')\nonumber\\&\times\sum_{l,m} G^{k'}_{lm}\delta_l I^{k'}_{lm},\\
    \frac{dR^k_{ij}}{dt} = &\, \gamma I^k_{ij}.
\end{align}

Vectorizing these equations using the method presented earlier, we obtain
\begin{align}
    \mathbf{J}=\frac{\beta}{\langle k\rangle}\mathbf{\hat{x}}\mathbf{\hat{y}}^T - \gamma \mathbb{I}
\end{align}
where
\begin{align*}
    \mathbf{\hat{x}} = \begin{bmatrix}k_1\mathbf{x} \\ k_2\mathbf{x} \\ \vdots \\ k_M\mathbf{x}\end{bmatrix}\!\!,\,
    \mathbf{\hat{y}} = \begin{bmatrix}k_1P(k_1)\mathbf{y_{k_1}} \\ k_2P(k_2)\mathbf{y_{k_2}} \\ \vdots \\ k_MP(k_M)\mathbf{y_{k_M}}\end{bmatrix}\!\!,    
\end{align*}
and $\mathbf{x}$ and $\mathbf{y_k}$ are defined as before.

Then, the largest eigenvalue is
\begin{align}
    \lambda &= \frac{\beta}{\langle k \rangle} \mathbf{\hat{y}}^T\mathbf{\hat{x}} - \gamma, \nonumber\\
    &= \frac{\beta}{\langle k\rangle}\sum^M_{i=1} k_i^2 P(k_i)\langle\delta\epsilon\rangle_i,
\end{align}
and the reproduction number is 
\begin{align}
    R_0 = \frac{\beta}{\gamma}\frac{1}{\langle k\rangle} \sum^M_{i=1} k_i^2 P(k_i)\langle\delta\epsilon\rangle_i.
\end{align}

This estimate closely matches the findings of Ref.~\cite{gou_how_2017}, the key difference being that this study instead examined individual heterogeneity in both infection and healing rates. When we assume $\langle\delta\epsilon\rangle_i=\langle\delta\epsilon\rangle$ for every $i$, we obtain
\begin{align}
    R_0 = \frac{\beta}{\gamma}\frac{\langle k^2\rangle}{\langle k\rangle}\langle\delta\epsilon\rangle,
\end{align}
which corroborates previous results for the network configuration model~\cite{pastor_satoress_2001}.

\section{Results}
We explore the effect of different correlations across communities on the behavior of an epidemic. We validate our mean-field approximations of the reproduction number against numerical simulations and demonstrate our theoretical results closely match those from simulation. We then use our theoretical $R_0$ value to describe how subgroup mixing structure affects the reproduction number. We then describe how community structure affects the temporal trajectories across different correlation structures. Lastly, we explore the impact of social intervention on each community and the entire population for a population comprising two communities. We use these cases to explore opportunities for intervention and control.

\subsection{\label{sec:validation} Numerical simulations validate analytical estimates of the reproduction number}
\begin{figure}[ht]
    \centering
    \includegraphics[width=\linewidth]{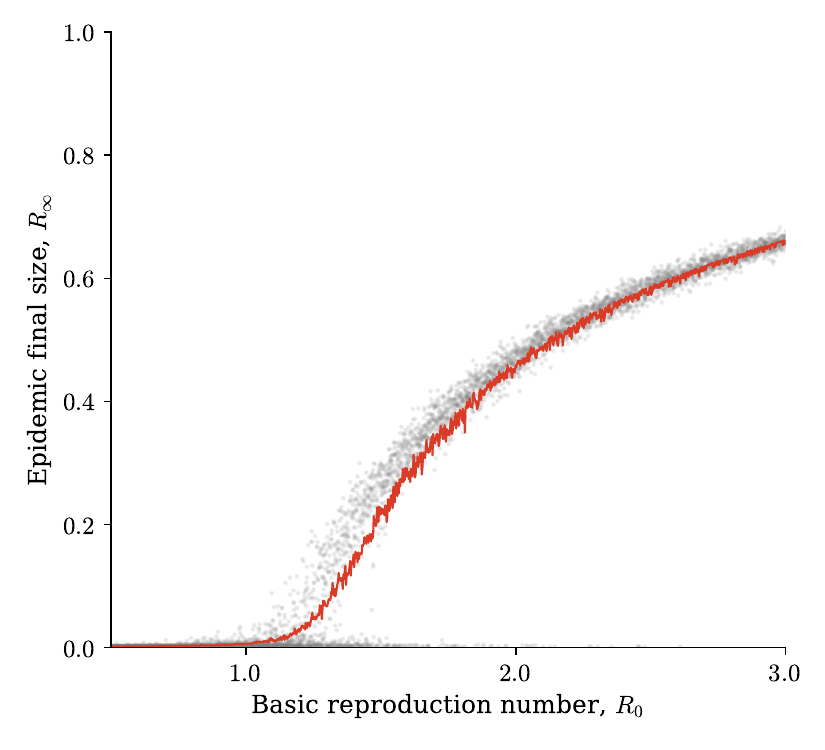}
    \caption{\textbf{Experimental validation of $R_0$ with numerical simulations.} Each gray point represents a single realization and the red solid line represents the mean of all realizations at that $R_0$ value. For this simulation, we used a network of size $N=10^4$ nodes with an mean degree of $10$. The community structure was specified by our imbalance parameter, $e=0.8$, and the fraction of population in community 1, $r=0.8$. We set our recovery rate, $\gamma=0.1$, and calculated the transmission rate, $\beta$, using our parameter values and the basic reproduction number, $R_0$.}
    \label{fig:numerical_validation}
\end{figure}

We validate our theoretical estimates of the basic reproduction number using stochastic agent-based simulations on networks. The joint distribution of susceptibility and transmissibility were modeled as a bivariate normal distribution and the correlations between the two were modified using the covariance matrix, $\Sigma$. The mean values of $\delta$ and $\epsilon$ were both $1.5$ and the variance of both was $1$. This distribution was then truncated so that it was supported on the region $[0, 3]\times [0,3]$ and the resulting distribution was used to calculate $\langle\delta\epsilon\rangle$. Then, individual values of $\delta$ and $\epsilon$ were sampled from this distribution, with different $\Sigma$ values for each community. 

Each network was sampled from the planted partition model with $10^4$ nodes and a mean degree of $10$. We considered imbalance parameters $e=0.8$ and community ratios $r=0.8$. We then represented these networks as weighted, directed networks with a reciprocity of $1$ and edge weights of $w_{ij}=\epsilon_i\delta_j$ for edge $(i,j)$.

We used the Epidemics on Networks (EoN) software~\cite{miller_eon_2019} to simulate each realization of heterogeneous SIR model using the \texttt{fast\_SIR} method, with $\gamma=0.1$, $\beta$ calculated based on the desired $R_0$ value, and transmission weights of $w_{ij}$. We initialized each simulation with $1\%$ of the whole population infected seeded uniformly at random. We considered $10^3$ values of $R_0$, equally spaced on the interval $[0.5, 3]$. For each $R_0$ value, we generated $100$ realizations of the planted partition model and ran an SIR simulation on each network. We calculated the epidemic final size as $R_\infty=R(t^*)$, where $t^*=\min(t\mid I(t)=0)$.

Fig.~\ref{fig:numerical_validation} demonstrates the accuracy of our theoretical results. Here, each data point is a single simulation, and the solid line represents the mean final size. We see that our theoretical estimates of $R_0=1$ accurately predict the empirical epidemic threshold. We present additional validation in Appendix \ref{sec:additional_validation}.

\subsection{The reproduction number depends nonlinearly on network structure and infection profiles}
\begin{figure*}[ht]
    \centering
    \includegraphics[width=\linewidth]{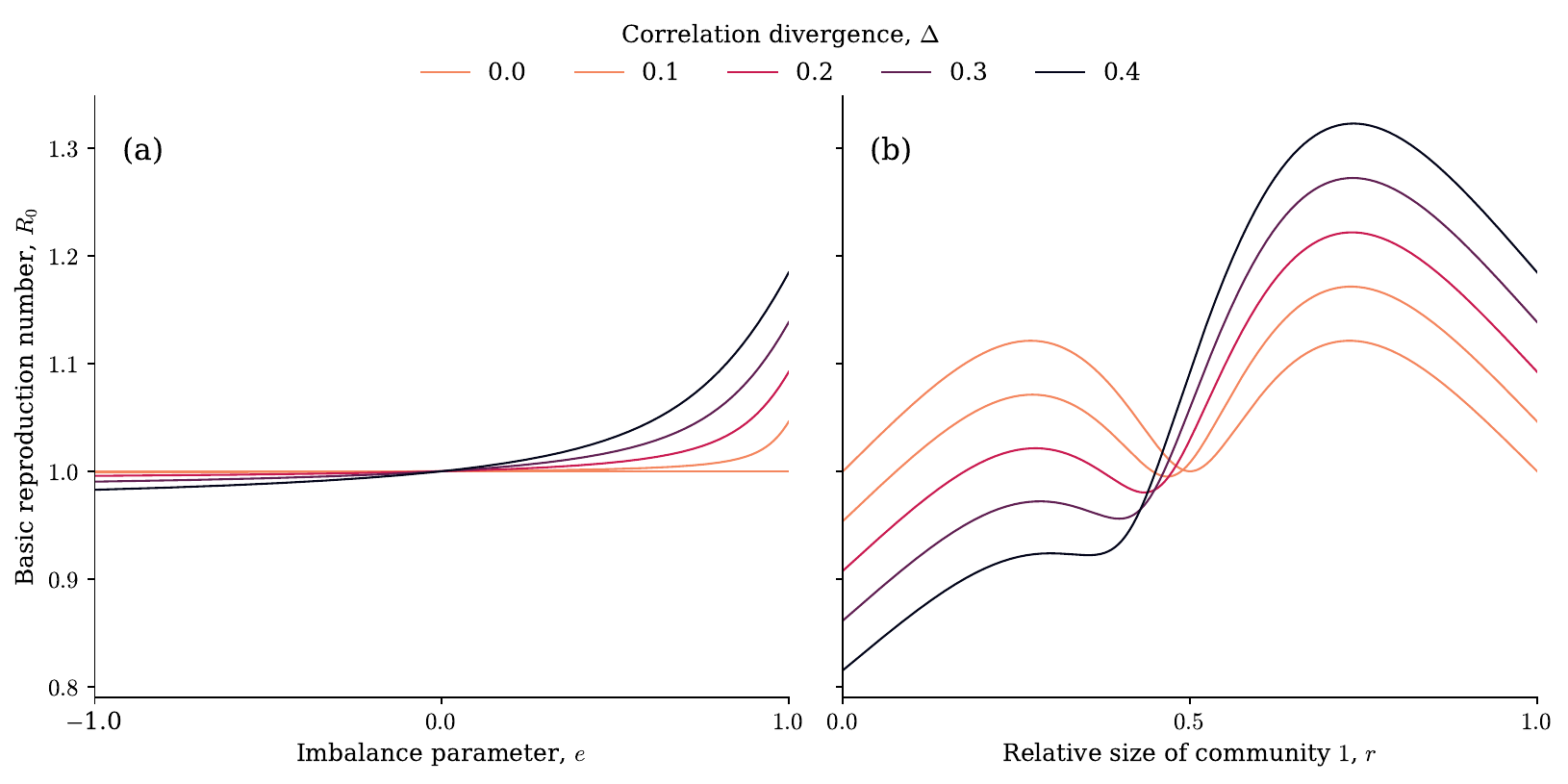}
    \caption{\textbf{Analytical estimates of the reproduction number, $R_0$, with respect to community strength and relative size.} Panel a) illustrates the effect of the imbalance parameter, $e$, on $R_0$, which we assign a value of $1$ in the uncorrelated case. Communities $1$ and $2$ have correlations between $\delta$ and $\epsilon$ of $-\Delta$ and $\Delta$, respectively. When $e=-1$, we obtain a bipartite network, $e=0$ corresponds to an Erd\H os-R\'enyi network, and $e=1$ creates two isolated communities. Panel b) illustrates the impact of imbalanced community sizes on $R_0$, where $r$ denotes the fraction of nodes in community 1.}
    \label{fig:r0_vs_r}
\end{figure*}

Having validated our analytical estimate of $R_0$, we quantify how $R_0$ is determined by both population structure and infection profiles within subpopulations. In Fig~\ref{fig:r0_vs_r}, we plot our theoretical predictions of $R_0$ with respect to, first, the imbalance parameter, and second, the relative sizes of the communities.

In the absence of transmission heterogeneity, Ref.~\cite{landry_opinion_2023} found that only when the sizes of the two communities are unequal does community structure affect $R_0$. Furthermore, this change in $R_0$ is symmetric around the balanced community case, i.e., $R_0(r) = R_0(1-r)$, where $R_0(r)$ denotes the reproduction number as a function of relative size of community 1. When adding correlated susceptibility and transmissibility that differ between communities, however, these two facts are no longer true. Without loss of generality, we let $\langle\delta\epsilon\rangle_1=\langle\delta\epsilon\rangle+\Delta$ and $\langle\delta\epsilon\rangle_1=\langle\delta\epsilon\rangle-\Delta$, calling $\Delta$ the \textit{correlation divergence}. We set $\beta$ and $\gamma$ such that when $\Delta=0$ and $r=0,0.5,1$, $R_0=1$.

In Fig.~\ref{fig:r0_vs_r}(a), we see that when $\Delta>0$, $R_0$ increases nonlinearly as the two communities become increasingly isolated. As described in Ref.~\cite{harris_infections_2025}, populations with larger correlations between susceptibility and transmissibility are associated with larger reproduction numbers. Therefore, placing links previously connecting two communities instead within a single community increases homophily between infectious traits; links connecting two individuals each with traits sampled from different distributions now connect individuals with traits sampled from the same distribution. This new structure makes it easier for an epidemic to spread in community $1$. Here, the intuition is that individuals with greater susceptibility are not only infected first~\cite{rose_population_2020}, but they are also more transmissible which, in turn, increases the likelihood of transmission to their neighbors. The pathogen will eventually spread in the other community as well once it is established in the positively correlated community. This effect increases with correlation divergence, and thus $R_0$.

Fixing the imbalance parameter and varying the relative sizes of the communities, we find that the effect on the reproduction number is asymmetric; when the community with more positive covariance between these traits is larger than the community with more negative covariance between traits, the reproduction number is larger than when the opposite is true. Similar to Ref.~\cite{landry_opinion_2023}, we observe that for both $r<0.5$ and $r>0.5$ there is a unique maximum value of $R_0$ and that the changes in $R_0$ are primarily driven by differential density across the network.

\subsection{Epidemic trajectories through subpopulations}

Having explored the effect of both structure and dynamics on the reproduction number, we now describe the temporal dynamics of our models in both communities. For all cases, we seed the epidemic in community $1$ with $0.1\%$ of the community infected. We consider strong community structure, $e=0.9$, and equal-sized communities, though our results match qualitatively in the case of unequally-sized communities. Here, we set the Pearson correlation coefficient between susceptibility and transmissibility for community $1$ to be $\pm 0.8$, with the opposite for community $2$, i.e., $\mp 0.8$.

Because Ref.~\cite{harris_infections_2025} demonstrated that positively correlated susceptibility and transmissibility in a population lead to faster transmission and larger infection peaks, while the opposite is true for negative correlations, we expect the same in a population comprising subgroups. Following this intuition, an epidemic always spreads first and reaches the largest epidemic final size in the positively correlated community. Furthermore, the difference between the final sizes of the two communities is much larger than the SIR model predicts. There is also asymmetry in the epidemic trajectories for the $(\delta,\epsilon)$-SIR model, where seeding an epidemic in the positively correlated community leads to higher simultaneous infection levels than are reached when seeding in the negatively correlated community. This asymmetry, however, is dwarfed by the asymmetry of dynamics between the positive and negatively correlated communities.

\begin{figure}[ht]
    \centering
    \includegraphics[width=\linewidth]{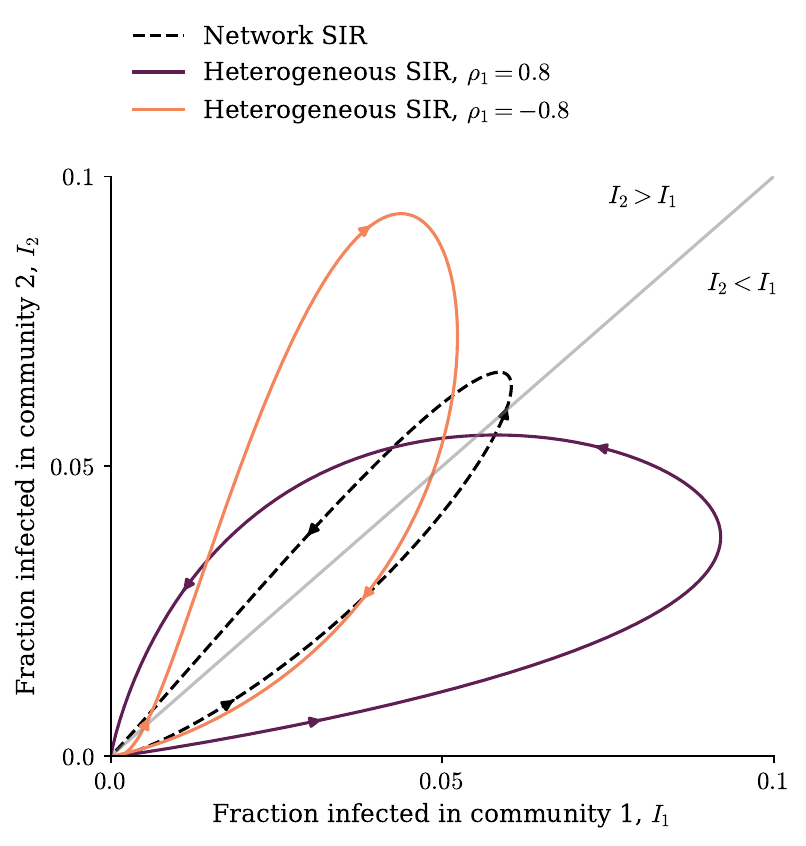}
    \caption{\textbf{Fraction of each community infected over time.} We plot the fraction of community $2$ infected with respect to the fraction of community $1$ infected, both as a function of time. For all cases, $e=0.9$, $\langle k\rangle=10$, $\gamma=0.1$, $R_0=1.5$, the two communities are of equal size, and $I_1=0.001$ and $I_2=0$ (with no recovered individuals). In all cases, the arrows denote the temporal direction on each curve. The black dashed line, the solid dark line, and the solid light line represent, respectively, the temporal dynamics of the SIR model, the $(\delta,\epsilon)$-SIR model with $\rho_1=0.8$ and $\rho_2=-0.8$, and the $(\delta,\epsilon)$-SIR model with $\rho_1=-0.8$ and $\rho_2=0.8$.}
    \label{fig:infection_dynamics}
\end{figure}

\subsection{Between-group social interventions can be costly for protecting vulnerable populations}
Having described the implications of heterogeneous transmission on epidemics, here we explore the effects of social interventions on social networks with community structure and community-specific joint distributions of susceptibility and transmissibility. One can easily imagine a young, healthy population in one community and an older, vulnerable population in another community, representing, for example, assisted living or nursing home settings. In this case, we might expect that for the young population, individuals who are more likely to transmit a disease are also most likely to contract it, and that the opposite is true for the elderly population.

When considering interventions in such a population, we quantify the impact of partially isolating these two communities. Because simply removing links between these communities reduces $R_0$, we add an additional constraint; namely, that the population has a constant mean degree. This constraint is equivalent to specifying a population-level social budget that must be satisfied; while we can trivially disconnect every node, this comes at a social cost.

\begin{figure}[ht]
    \centering
    \includegraphics[width=\linewidth]{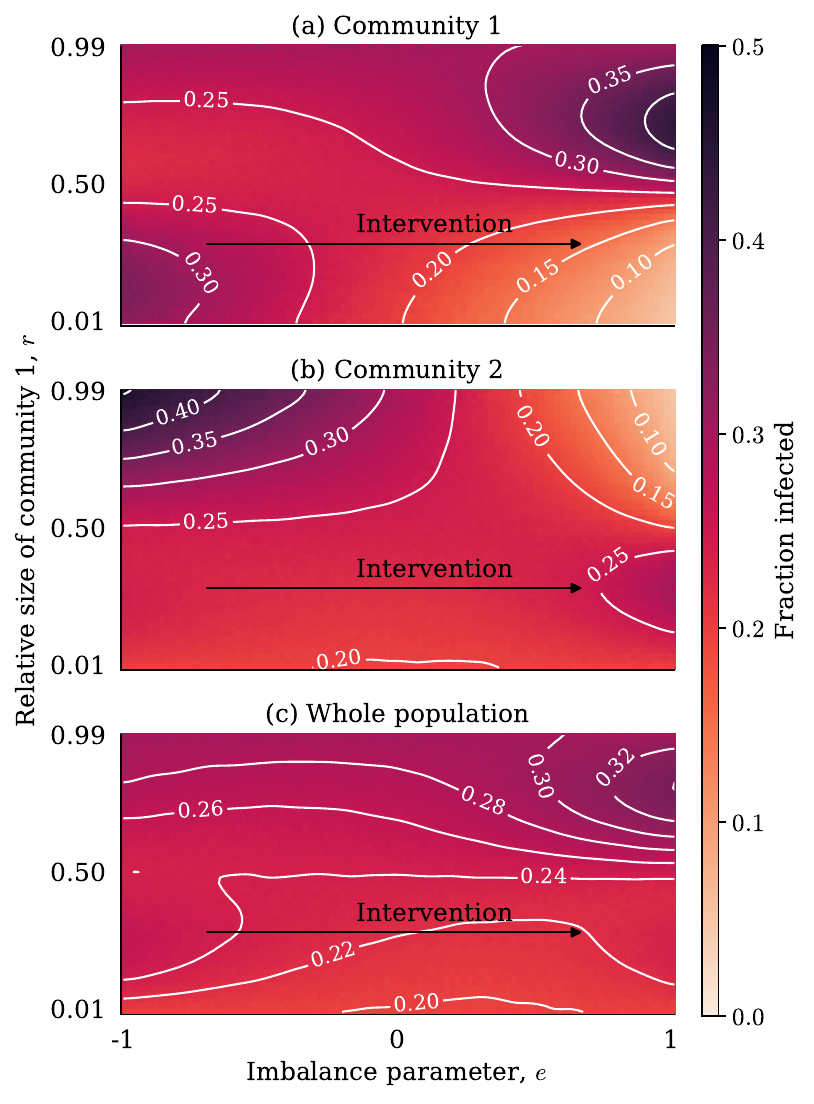}
    \caption{\textbf{The effect of interventions under a fixed social budget.} Here, we consider a planted partition model with $10^4$ nodes and a mean degree of $10$. Communities $1$ and $2$ have $\delta, \epsilon$ correlations of $0.8$ and $-0.8$ respectively. We simulate the spread of a contagion with $5\%$ of the nodes in the whole population infected initially. We set our recovery rate, $\gamma=0.1$, and calculated the transmission rate, $\beta$, assuming $e=0$, $r=0.5$, and the basic reproduction number, $R_0=1.1$. The black arrow shows an intervention strategy where reducing the contacts between communities can reduce the infection size.}
    \label{fig:heatmap_g2}
\end{figure}

We specifically wanted to examine cases where the reproduction number is close to $1$, and $R_0$ for communities $1$ and $2$ in isolation are above and below $1$ respectively. Does isolating these communities protect vulnerable subpopulations? Fig.~\ref{fig:heatmap_g2} illustrates the epidemic final sizes in communities $1$ and $2$ and the entire population. Fig.~\ref{fig:heatmap_g2}(a) displays the fraction of community $1$ (the correlated subpopulation) infected for different strengths of community structure, $e$, and different relative community sizes, $r$. The epidemic final size is largest for large $r$, large $e$, corroborating our results in Fig.~\ref{fig:r0_vs_r}(b) for the reproduction number. This can be understood as a combination of increased mean degree in community $1$, and increased concordant links due to between-community edges being placed within the community. Fig.~\ref{fig:heatmap_g2}(b) shows quite different trends; for large $r$ and $e$, transmission is lowest, indicating that while it is possible to protect this community, it comes at cost of infections in the other community. For the entire population, high $e$ and $r$ actually \textit{increases} the total number of infections, indicating that this intervention not only increases the number of infections in community $1$, but increases the total number of infections in the entire population as well. Our results are a cautionary tale against isolation-based network interventions.

\section{Discussion}
In this paper, we demonstrated how heterogeneity in both infectious traits and social connectedness can together shape how an epidemic spreads. In particular, we derived the basic reproduction number for a population with two subgroups and group-specific joint distributions of transmissibility and susceptibility. We not only show that the reproduction number nonlinearly increases as these two communities become increasingly separated but also that the reproduction number is sensitive to the interaction between the relative sizes of the communities and the distribution of their infectious traits. In particular, the reproduction number is maximized when the community with the most positive correlation between its infectious traits is larger than half the population. Our derivation of $R_0$ for the configuration model is similar to previous estimates but now, the second moment is weighed by the correlation between the susceptibility and transmissibility for each degree. We validated these analytical solutions with numerical simulations.

These findings will help policymakers design effective interventions for pathogens characterized by differential susceptibility and transmissibility across individuals. First, we showed that similar to the findings of Ref.~\cite{harris_infections_2025}, that the temporal behavior of epidemic models incorporating heterogeneous infectious traits differs significantly compared to epidemic models without trait heterogeneity. In addition, for a population with two communities, each with its own distribution of infectious traits, these temporal differences are asymmetric based on the community to which patient zero belongs.

Lastly, the effect of social interventions in a social network with two subpopulations, when trying to protect one of the subpopulations largely depends on the community's relative sizes and the difference in its infectious traits. If the size of the subpopulation we wish to protect contains fewer than half the population, then removing links between subpopulations as an intervention strategy efficiently reduces the size of the epidemic in this population.

Age-structured epidemic models have been heavily utilized when determining vaccination strategies~\cite{bubar_modelinformed_2021}, forecasting~\cite{buchwald_estimating_2021}, and understanding the differential roles of age groups in the spread of an epidemic, because of their ability to model age-specific risk, both in infection and transmission. These models are informed by large-scale studies of age-mixing structure~\cite{prem_projecting_2017, mossong_social_2008}, construction of synthetic social networks from age-mixing matrices~\cite{mistry_inferring_2021}, and estimations of age-specific susceptibility and infectiousness~\cite{franco_inferring_2022}. These models, however, rarely consider both susceptibility and transmissibility as correlated traits, and incorporating this feature in models used to make public health decisions is an opportunity for future work.

A limitation of our work, however, is that our model is not data-informed. While several studies have estimated the variation in susceptibility and transmissibility independently, these traits are likely correlated and to our knowledge, no studies have reconstructed their joint distribution without this independence assumption. However, Ref.~\cite{sender_unmitigated_2022} inferred the joint distribution of the generation interval, which is a property of the infecting individual, and the incubation period, which is a property of the infected individual. Similarly, Ref~\cite{anderson_quantifying_2023} inferred the heterogeneity in infectiousness and susceptibility, however this study was constrained by household size and also estimates these quantities independently. Additional studies estimating the joint distribution of susceptibility and transmissibility for different demographics will improve the utility of our model. Knowing the underlying contact network, and infection data from multiple outbreaks, we may be able to leverage inferential approaches (for example, Refs.~\cite{landry_reconstructing_2024, peixoto_network_2019}) to estimate these joint distributions.

While in this study we considered two communities for analytical tractability, one could, in principle, model any number of communities to more closely model empirical population structure. In addition, we could also combine our analysis for the stochastic block model with our analysis for the configuration model to understand the epidemic dynamics on a degree-connected stochastic block model~\cite{karrer_stochastic_2011}. There are also opportunities to model biologically-informed interactions between susceptibility and transmissibility. While in our study susceptibility and transmissibility may be correlated, we assumed that their \textit{effects} are independent, i.e., the probability of an individual $i$ infecting individual $j$ is $\beta\delta_i\epsilon_j$, and one can describe nonlinear interactions between $\delta$ and $\epsilon$ with a general function $T(\delta,\epsilon)$, similar to the framework proposed in Ref.~\cite{miller_epidemic_2007}.

Nonetheless, our framework provides a starting point for analyzing heterogeneity in both infectious traits and contact structure. We have demonstrated that both play important roles in the spread of epidemics.

\section*{Data availability}
The code and synthetic data supporting this work are openly
available on \href{https://github.com/AbhayGupta115/HeterogenousCommunities}{GitHub} and at Ref.~\cite{gupta_code_2026}.
\vspace{0.2in}
\section*{Acknowledgements}
N.W.L. and A.G. acknowledge support from the University of Virginia Prominence-to-Preeminence (P2PE) STEM Targeted Initiatives Fund, SIF176A Contagion Science. N.W.L. and A.G. acknowledge helpful feedback from Laurent H\'ebert-Dufresne and Daniel Kaiser.

\bibliography{references}

\appendix
\section{\label{sec:additional_validation} Additional validation of our analytical results}
\begin{figure*}[ht]
    \centering
    \includegraphics[width=\linewidth]{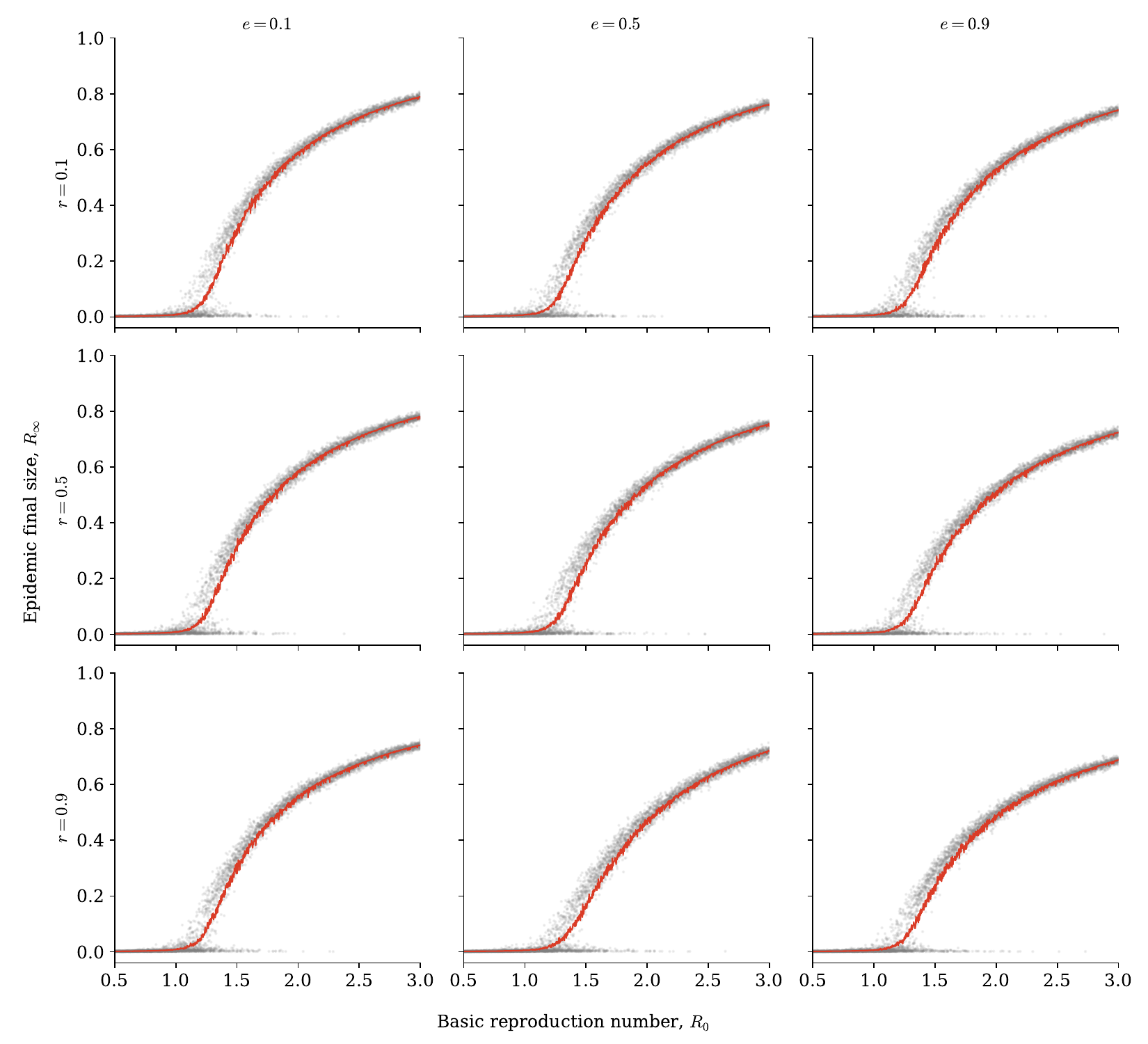}
    \caption{\textbf{Experimental validation of $R_0$ with numerical simulations.} Each gray point represents a single realization and the red solid line represents the mean of all realizations for a particular $R_0$ value. Each simulation was run on a planted partition network with $10^4$ nodes with a mean degree of $10$. The planted partition model is characterized by the imbalance parameter, $e$, specifying the fraction of edges connecting two individuals from the same community, and $r$, specifying the fraction of nodes in community $1$. Each row corresponds to a fixed value of $r$ and each column corresponds to a fixed value of $e$.}
    \label{fig:grid_validate}
\end{figure*}

\vspace{-0.05in}
To validate our analytical results across a range of community structures, we consider imbalance parameters of $e=0.1, 0.5, 0.9$ and community relative sizes of $r=0.1, 0.5, 0.9$. We use the same epidemiological parameters as described in Sec.~\ref{sec:validation}. In Fig.~\ref{fig:grid_validate}, we see strong agreement with our analytical predictions across a wide variety of structural parameters.

\end{document}